# Multi-line Doppler imaging of MR Ser in high-state

M. P. Diaz[1] and D. Cieslinski[2]

[1] IAG, Universidade de São Paulo, 05508-900, São Paulo, SP, Brazil
marcos@astro.iag.usp.br

[2] Instituto Nacional de Pesquisas Espaciais, INPE, 12227-10, São José dos Campos, SP, Brazil
deo@das.inpe.br

ABSTRACT

Doppler images in Balmer, HeI, HeII and CII lines, and simultaneous I-band photometry of the polar MR Ser are presented and analyzed. The Balmer and Helium Doppler tomograms of this bright polar at high mass transfer state show the emission from the accretion flow and the heated surface of the companion star. As a result of a comparison between the Doppler tomograms the ionization structure of the flow could be constrained. The highest ionization region was found in the vicinity of the magnetospheric radius. Photoionization modeling of the accretion column indicates that the Balmer and Helium emission line production in this system can be explained by the central soft X-ray illumination only. The orbital ephemeris of MR Ser has been revised.

## 1. INTRODUCTION

Polars or AM Her stars are cataclysmic binaries that harbor a high field ($10^7$ to $10^8$ G) magnetic white dwarf or primary (see Warner 1995 for a review). In these systems the magnetic pressure becomes larger than the ram pressure of the infalling gas inside the binary gravitational potential at large distances from the primary, preventing the formation of an accretion disk. Instead, the gas flows supersonically from a free-particle path to a magnetically funneled accretion trajectory in the magnetosphere of the white dwarf. The velocity of the gas increases as it follows the field lines towards the white dwarf. A steady shock is formed close to the white dwarf surface with the post-shock gas being cooled mainly by thermal bremsstrahlung X-ray emission with rates that are close to the accretion luminosity ($10^{33}$ to $10^{34}$ erg.s$^{-1}$ in high states). Some of this emission is absorbed by the primary where it thermalises and is re-emitted into the X-rays / EUV bands. Depending on the white dwarf field strength a different balance between hard and soft X-rays is expected (Beuermann & Schwope 1994). Illumination of the secondary star by this bright ionizing source is often seen in polars during their high mass transfer states, both in the continuum and in recombination lines. In this paper we study the effects of this high energy source on the accretion flow itself.

The typical optical emission line spectra of a polar in high mass transfer state shows flat Balmer decrements, HeI lines and strong HeII lines. The NIII + CIII complex at 4640-4650 Å is also seen. In particular, the presence of the OIII 3429 Å in MR Ser (PG1550+19) indicate a possible contribution from Bowen fluorescence to NIII 4640 Å and other NIII lines (Williams & Ferguson 1983). The accretion stream and column as well as the heated surface of the Roche lobe filling donor star are known to be prominent contributors to the emission line flux. However, the identification of heating sources and line production processes in the accretion flow of magnetic CVs is still model dependent. By analyzing integrated line fluxes of the UV resonant lines (CIV 1550 Å, CII 1335 Å, SiIV 1400 Å, and NV 1240 Å) and HeII 1640 Å from several polars, Howell et al. (1999) suggested that these lines may be produced in a photoionized gas. On the other hand, Ferrario & Wehrse (1999) showed that the integrated HeII to Balmer line ratios seen in polar spectra can only be reproduced by models that consider both X-ray heating and magnetic reconnection dissipation close to the coupling radius $R_c$.

The polar MR Ser is a classical magnetic CV (PG1550+191) found in the Palomar Green blue objects survey (Liebert et al. 1982). It is a well studied AM Her type variable with known system (Mukai & Charles 1987; Shahbaz & Wood 1996) and magnetic parameters (Wickramasinghe et al. 1991; Schwope et al. 1993). K-band spectroscopy shows a M8V secondary with normal Carbon abundances (Harrison, Osborne & Howell 2005). Distance estimates (Araujo-Betancor et al. 2005, Schwope et al. 1993) and X-ray observations while the system was in low- (Ramsay et al. 2004) and high-state (Angelini, Osborne & Stella 1990) are also available. The reddening to this system was considered negligible in the optical, following the absence of any interstellar signature at 2200 Å (Szkody, Liebert & Panek, 1985).  Shahbaz & Wood (1996) presented Doppler maps of the CaII 8498, 8542 Å emission claiming that this triplet is formed close to the inner Lagrangian point (L1). Their data were possibly taken while the system was fading to or at a low photometric state. MR Ser stayed predominantly at low mass transfer states from 2001 to 2004 (Kafka & Honeycutt 2005) with corresponding bolometric luminosities of ~$10^{30}$ erg.s$^{-1}$ (Ramsay et al. 2004). The system was found in a high photometric state in 2006 and 2007 (this work) allowing a phase-resolved study of its emission line profiles during high mass transfer rates. In the next section we describe the spectroscopic and photometric observations of MR Ser. In sections 3 and 4 the Doppler imaging results and a photoionization modeling of the accretion flow are presented and discussed.

## 2. OBSERVATIONS

### 2.1 *Spectrophotometry*

The observations of MR Ser analyzed in the present work were obtained at Laboratório Nacional de Astrofísica - LNA at Brasópolis, Brazil in 4 observing runs during 2006 and 2007 (table 1). Time-resolved differential spectrophotometry was obtained at the 1.6 m telescope using a Cassegrain spectrograph with a spectral resolution of 1.9 Å (FWHM) or ~120 km/s. The spectral *PSF* is oversampled by 3.5 pixels (FWHM). Individual target exposures comprises an orbital phase interval of ~0.07. A total of 304 spectra sampling ~20 orbital cycles were taken, covering from 4070 Å to 5010 Å. A comparison star was included in the slit at *PA*=286 degrees and 78 arcsec away from MR Ser in order to correct for small transparency changes during the exposures as well as slit losses and atmospheric dispersion effects in first order. Tertiary standard calibration stars from the list of Hamuy et al. (1994) were observed employing wide slits widths under photometric conditions. Target exposures were frequently bracketed by He-Ar frames yielding a continuous wavelength calibration with *RMS* < 5 km/s. Twilight and dome flats were also taken on a daily basis. The data reduction using IRAF[1] packages included standard bias, flatfield, illumination correction procedures and optimal extraction of spectra, followed by wavelength and flux calibration. The total spectrophotometric calibration errors are estimated to be <0.12 magnitudes while typical continuum *S/N* ratio of individual spectra is ~10.

---

[1] IRAF is distributed by the National Optical Astronomy Observatory, which is operated by the Association of Universities for Research in Astronomy, Inc., under cooperative agreement with the National Science Foundation.

## 2.2 *I-band Photometry*

Simultaneous photometric observations of MR Ser were obtained on the night of 30 May, 2006 (UT) with the 0.6-m Zeiss telescope of LNA. The data were collected using a Wright Instruments thermoelectrically cooled camera with a EEV CCD02-06-1-206 (385x578 pixels) thin and back illuminated chip, which was operated in the frame-transfer mode. The observations were done in the Kron-Cousins I-band using the fast photometer mode, with an integration time of 30 seconds and dead time of a few ms between integrations. Timing for the instrument is provided by a global position system (GPS) receiver. The data were reduced using the aperture-photometry routines of the IRAF APPHOT package. The images were debiased and corrected for flat-fielding. Fluxes for the variable star and for four field stars (used as comparisons) were then extracted using a circular aperture of 4 arc seconds of radius. An estimate of the uncertainty in the differential measurements of ~0.02 magnitudes can be obtained from the dispersion in the differential photometry of similar brightness, constant stars used as comparisons.

## 3. RESULTS

### 3.1 *Orbital Ephemeris and I-band Light Curve*

The orbital period of MR Ser was first determined when the object was identified as a AM Her type CV by Liebert et al. (1982). These authors used the linear polarization pulses as timings of orbital phase in this synchronous polar. A long-term spectroscopic ephemeris for the secondary star was derived by Schwope et al. (1991, 1993) making use of the sinusoidal radial velocity curves of the narrow emission line component in Balmer and HeII 4686 Å lines and the NaI 8183, 8194 Å absorption lines. They also confirmed the previous finding by Liebert et al. (1982) that the inferior conjunction of the companion star corresponds to the minimum in the V-band light curve. Using RoboScope V-band data in high-state taken for 13.3 years (1991-2004) Kafka & Honeycutt (2005) found that minimum light had drifted to phase 0.82 using the Schwope et al. (1993) ephemeris, suggesting the need to revise the orbital ephemeris. The ephemeris given by Schwope et al. (1993) is consistent, within the uncertainties, with the ephemeris given by Shahbaz & Wood (1996) which was derived using data taken in 1994, though the period derived by Schwope et al. (1993) is the more accurate and good enough to maintain cycle count throughout our data set.

In the present work we have measured the narrow component of Balmer lines in 2006 and 2007 finding that the negative-to-positive crossing of the radial velocity curve takes place at phase 0.72, thus confirming the need of an updated spectroscopic ephemeris. The cycle-counting of our radial velocity data can be performed using the period from Schwope et al. (1993). We have verified that such a value is certainly precise enough to perform the Doppler tomography analysis of our dataset, which spawn for only 390 days. The secondary conjunction phase was revised using the negative-to-positive crossing of H$\beta$ and H$\gamma$ narrow component, yielding the following ephemeris (where the original period by Schwope et al. (1993) and its uncertainty have been retained):

$T_0$ = HJD2453885.3466(14) + 0.07879793(8) $E$ \hfill (1)

Simultaneous I-band photometry was phase-folded with the ephemeris above yielding a minimum at spectroscopic phase 0.0 (figure 1). A comparison between the phase of minimum light in our I-band light curve and the V-band results from Kafka & Honeycutt (2005) indicates that the accretion region responsible for the continuum emission has not shifted in phase once the correct ephemeris is applied. The light curves in both V- and B-bands are not sinusoidal and may be interpreted as the varying aspect of the continuum sources in the system and the contribution from the cyclotron harmonics in the optical (Schwope et al. 1993). In addition, one may expect that a secondary which develops a bright chromosphere will also increase its photospheric temperature because of the illumination by soft X-rays. This component may contribute to the rounded shape of the maximum seen at phase ~0.6 in the I-band light curve, although there is a phase shift of 0.1 that could not be explained in a simple illumination scenario. By subtracting a simple sinusoid from the orbital light curve we find a second modulation with half the orbital period. Its amplitude is inconsistent with an ellipsoidal modulation of the companion, which is unexpected to be seen in high state. A periodogram of the lower panel data was computed showing a significant (8 - 10 sigma) peak at Porb/2. Although it may sound convincing, a much larger number of cycles is required to study the phase stability of this residual modulation. The phase of maxima of this residual light curve $\phi_{ORB}$(max) ~0.2 and ~0.7 and the presence of two cycles per orbit suggest that such modulation may be due to the changing aspect of the optically thick part of the accretion column combined with the cyclotron beaming effect. However, a detailed modeling of the continuum emission regions is required to confirm this interpretation.

### 3.2 *Phase-resolved Spectra and Doppler Tomography*

The emission line profiles in MR Ser show a structured and complex orbital behavior. Both Helium and Balmer lines present a narrow component with sinusoidal velocity curves that were previously observed and recognized as the emission from the heated surface of the red dwarf (Schwope et al. 1991). Measurements of the Balmer line flux show a maximum around $\phi_{ORB} = 0.55$ or close to the maximum aspect of the illuminated surface of the secondary. This phase is also near the I-band continuum maximum. The maximum HeII/H$\beta$ ratio takes place at $\phi_{ORB} \sim 1.05$ (figure 2). The overall shape of the lines is defined by the orbital phase with stochastic variability on top of that. The similarity between the Doppler maps from 2006 and 2007 data (see next session) indicates that a secular component, if present along that year, was small. We have estimated the non-orbital variability by computing the *RMS* line profile variation about the average profile in phase bins. The rms/flux variations found over the H$\beta$ line core are below ~0.25. However, a quantitative evaluation of non-orbital behavior would require a much larger data set. The trailed spectrograms of the H$\beta$ and HeII 4686 Å continuum subtracted lines show profiles that vary in shape, width and intensity (figure 3). These trailed spectrograms were calculated by phase-binning the entire dataset. Around $\phi_{ORB} \sim 0.45$ the HeII line presents a double peak, while the H$\beta$ line displays an asymetric profile with a red peak.

    The Doppler imaging of the emission lines (Marsh & Horne 1988) in magnetic CVs has been proved as an useful technique for locating and measuring the line emissivity of the accretion flow and companion star in those systems (Schwope 2001). Doppler maps of Balmer, HeI, HeII and CII lines of MR Ser in high mass transfer state (figure 4) were calculated using continuum subtracted and flux calibrated line profiles. All available spectra without any phase-binning were used in these reconstructions. The filtered back-projection algorithm (Rosenfeld & Kak 1982) was employed to compute the velocity maps using only the observed line profiles as constraints. The usual coordinate system definition is adopted in our maps; i.e., the origin is at the binary rest, the *X*-axis points from the primary to the secondary star, while the *Y*-axis points in the direction of the secondary orbital motion. The velocity scale is corrected by an orbital inclination of 45 degrees (Brainerd & Lamb 1985). Typical

resolution of our maps is ~130 km/s (FWHM) or close to the instrumental limit. A self-consistency check was applied to all maps by comparing the line profiles derived from the reconstructions with the input profiles. Mapping of 3D structures, like the gas raising from the orbital plane near the coupling radius, using 2D data presents additional problems for all Doppler tomography algorithms. In the absence of additional observational input to the reconstruction the imaging of those 3D structures is smeared by the differences between the binary systemic velocity and gamma velocities of the gas in the column (Diaz & Steiner 1994). The expected amount of "gamma smearing" or intrinsic PSF of the accretion column emission was computed along the accretion column within the region sampled by the tomograms. The column path is plotted assuming a dipole surface field of 28 MG, derived from the photospheric Zeeman features seen in low state, and the field colatitude and longitude $(\beta,\varphi)$ given by Cropper (1988) and Schwope et al. (1993). The ballistic part of the flow is calculated assuming a mass ratio $q = M_2 / M_1 \sim 0.12$ with $M_1 \sim 0.6\ M_{SUN}$ and $M_2 \sim 0.07\ M_{SUN}$ (Mukai & Charles, 1987 and Shahbaz & Wood, 1996). Uncertainties in the white dwarf mass are very large. The value adopted here is between $0.5\ M_{SUN}$, as given by Schwope et al. (2000), and the value of $0.67\ M_{SUN}$ found by Shahbaz & Wood (1996). A value for the velocity at the magnetic coupling radius is marked on Doppler maps as a reference for $M\text{dot} = 3\times10^{-10}\ M_{SUN}/yr$ and a surface dipole footprint fractional area $f \sim 0.001$. Such a mass transfer rate is well within the accretion luminosity range allowed by the X-ray flux measurements at high state made by Angelini, Osborne & Stella (1990). The ratio between hard and soft X-rays fluxes $F_{BREMS} / F_{BB}$ may be as low as 0.1 for a magnetic field intensity $B = 28$ MG (Beuerman & Schwope 1994).

In the case of MR Ser the gamma smearing is not significant in the mapped flow and would not be enough to prevent the detection of the innermost part ($r < 10\ R_{WD}$) of the accretion column, which is not seen in any of the reconstructions. The PSF degradation of the column due to gamma smearing in this region is less than 200 km/s, reaching a maximum value of 300 km/s at $r \sim 5$ $R_{WD}$ in our maps. Enhanced emission from a region close to L1 is detected in all mapped transitions and, in contrast with some other polars (e.g. QQ Vul; Schwope et al. 2000), the ballistic stream emerging from $L_1$ is not seen as a well defined structure. Instead, asymmetric emission appears at higher $V_y$ velocities. This feature may be the signature of the gas coupling to the field lines, initially increasing its $V_y$ velocity as it leaves the orbital plane and align with the curved dipole field lines. However, higher resolution maps are required to confirm this hypothesis. It is evident from the maps that the Balmer lines are less concentrated and also fill a larger volume in velocity coordinates than the high ionization lines. Conspicuous differences between Balmer, HeI and HeII maps are also found suggesting that the ionization varies along the accretion flow. The faint CII 4267 Å line behaves much like HeII 4686 Å. The former has a very high upper level excitation potential (21 eV) and it is more likely produced by CIII recombination instead of collisional excitation. Self-absorption effects are more noticeable in the HeI triplet lines than in the singlet transitions, yet no significant differences could be found between the HeI 4922 Å and HeI 4471 Å reconstructions. If just photoionization is considered the HeII, HeI, CII and Balmer lines require photons with energies in excess of 54.4, 24.6, 24.4 and 13.6 eV, respectively.

The local HeII 4686 Å to Hβ line ratio (figure 5) may be used to probe the ionization structure. Although the inner face of companion star seems to be the brightest HeII source in the system, we conclude from the line ratio map that neither the ballistic stream nor the secondary star are the most ionized regions. Instead, the higher velocity gas in the vicinity of the coupling region displays extreme HeII to Hβ ratios, reaching ~1.7. Much lower values around 0.6 are seen at the expected locus of the low velocity ballistic stream. Average values near ~1.0 are found close to L1. Unfortunately, the instrumental noise is amplified in those rational maps as the low emissivity regions in the Hβ map are also the noisiest. This effect prevented the analysis of small-scale structures in the other line ratio maps.

However, large sections in the intensity maps were averaged yielding reliable measurements of each line emissivity around the coupling region. The measured line emissivity ratios are presented in table 2. Doppler maps of H$\beta$ and HeII using data subsets from each observing season were computed and measured separately showing similar results.

### 3.3 *X-Ray Heating Models*

The Doppler reconstructions combined with a kinematic model of the flow may be employed to derive local emission line ratios. These ratios on their turn can be used to constrain the physical conditions of the gas. The kinematic model adopted here is based on the assumptions of conservative mass transfer, zero-viscosity ionized gas and a dipole field geometry. Using those conditions and the Roche potential we may express the velocity vector as a function of the position inside the primary Roche lobe for both the free-particle stream and the accretion column (Ferrario & Wehrse 1999). The material couples to the field lines in a region with dimensions $r\Delta\phi\Delta r$, where $r$ and $\phi$ are spherical coordinates centered at the white dwarf. The magnetic field parameters used in previous sections and a range of mass-transfer rates were explored to estimate the Hydrogen density along the accretion column assuming solar abundances. All density profiles present small gradients at large distances where the flow shrinking imposed by the magnetic field is almost compensated by the increase in velocity. The mass transfer rate is basically unknown with the only effective constraint given by the X-ray observations in high-state by Angelini, Osborne & Stella (1990). The energy of the X-ray ionizing source is also constrained by these authors to $6 < kT_{BB} < 53$ eV. Given the dipole direction in this system the column, coupling region and ballistic stream are exposed to the bright soft X-ray emission from the accretion spot at the white dwarf surface. Typical mass density values along the column of MR Ser, calculated using $\Delta\phi = 25$ degrees and $\Delta r = 1$ $R_{WD}$ (e. g. Ferrario & Wehrse 1999), are shown as a function of $r$ in figure 6. An average between the distance estimates of 160 +/- 18/26 pc by Araujo-Betancor et al. (2005) and 139 +/- 13 pc by Schwope et al. (1993) is adopted. Calculations were performed with version 06.02.09b of CLOUDY, last described by Ferland et al. (1998) and a subplex algorithm optimization module. Our simulations estimate the local physical conditions and line emissivity of the upper accretion column illuminated from outside by the central soft X-ray source at the white dwarf. Radiative transfer of diffuse and central source ionizing radiation along the column is neglected. Line optical depth and collisional effects are important for both Hydrogen and Helium lines and were taken into account. A large grid of models was computed aiming to reproduce the average line ratios observed in the vicinity of the coupling region. The average Helium to Balmer ratios, Balmer decrements as well as the absolute line fluxes near the coupling radius are well reproduced by a model with $N_H$ near $10^{13}$ cm$^{-3}$ at the column (or $M$dot $\sim 10^{-10}$ $M_{SUN}$/yr) , $L_X = 10^{33}$ erg.s and $kT_{BB} = 12$ eV. Sample model values are given in table 2. These best fit parameters for the central source are consistent with the EXOSAT observations and expected interstellar Hydrogen column densities (Angelini, Osborne & Stella 1990). The X-ray source flux was constrained to be lower than the accretion luminosity for any given $M$dot. However, given the number of degrees of freedom and the chi-square topology, it is not possible to claim that the best-fit solution found is unique. Significant variations in $N_H$ can be compensated by changes in the ionization parameter and distance. However, there is a robust, if not unique, solution that matches the data using a source flux consistent with the limits on the accretion luminosity. In agreement with the HeII/Hβ map, model predictions suggest that HeII emissivity relative to Hβ decreases towards the secondary, well before reaching L1 (figure 7). High electron temperatures can be achieved by photoionization heating specially at outer parts of the flow.

On the other hand, depending on the input parameters a neutral Hydrogen zone can be formed inside the column if no other source of heating is claimed. Models were iterated to correct for line self-absorption effects but model fluxes are still less acurate at Hydrogen densities $N_H$ around or above $10^{14}$ cm$^{-3}$. Therefore, the lower parts of the column ($r < 7$ R$_{WD}$) could not be reliably modeled. We have not attempted to model the emission at the illuminated companion star surface.

## 4. DISCUSSION

The Doppler tomograms of the recombination lines in polars often show a conspicous signature of the free stream. All MR Ser Doppler maps share the absence of a well defined ballistic gas stream from L1, a common feature in tomograms of polars at high mass transfer states. There are other examples of systems with faint or unseen free-streams like the asynchronous polar BY Cam, which also lacks a bright focused stream from L1 in their Doppler reconstructions (Schwarz et al. 2005). In BY Cam the emission spread over a large velocity space due to the proposed accretion curtain scenario for this asynchronous system. The H$\alpha$ and HeI Doppler maps of BL Hyi (Mennickent, Diaz & Arenas 1999) show a featureless emissivity distribution but also lack a well defined ballistic stream. In the case of MR Ser a relatively low plasma emissivity at the free flow seems to be the cause of the hidden stream. We have compared HeII and H$\beta$ tomograms calculated using only the first or second year data. By using these two data sets it is not possible to claim the presence of significant differences that would indicate a change in the accretion geometry during high-state.

The photoionization/heating models presented in section 3.3 may be improved by a rigorous treatment of the radiative transfer along the accretion column. Nonetheless, the existence of a stable solution indicates that the emission lines from such a region can be produced, at least in MR Ser during high-state, by photoionization alone, although the presence of an additional weak source of heating like magnetic reconnection can not be completely ruled out considering the uncertainties in our model input parameters. Magnetic reconnection and dissipation processes in other stellar scenarios are usually short lived or bursting phenomena. In polars, magnetic dissipation near the coupling radius (Ferrario & Wehrse 1999) may eventually contribute to the production of emission line flickering seen in UV emission lines (e. g. Greeley et al. 1999). Considering that the recombination time-scales of the plasma near the coupling radius is of the order of few seconds one may expect that the release of magnetic energy via emission lines would be a local phenomenon. In this scenario, Doppler imaging of the emission line flickering sources would be an observational tool for studying the role of magnetic heating in polars, though this is beyond the scope of the data presented here. By using the Balmer decrements as an estimate for the gas temperature, the line forming regions in other polars were also found to be intrinsically hot, with temperatures between 10000 K and 15000 K (Gerke, Howell & Walter 2006). Even hotter conditions should prevail in the lower column, decreasing the Balmer line emissivity as suggested by the Doppler reconstructions.

## 5. CONCLUSIONS

Doppler images in the optical recombination lines of this bright polar were used to constrain the line emissivity distribution at high state. The line emission from the free gas stream close to L1 is absent or too faint to be detected as an individual structure. A bright Balmer and Helium line source is seen at (or near) the inner face of the secondary star while the I-band light curve suggests the presence of an optically thick continuum source at the accretion column. The highest ionization was found in the vicinity of the magnetospheric radius as a well defined structure in the HeII/Hβ map. Photoionization modeling of the accretion column indicates that the Balmer and Helium emission line production in the accretion flow can be explained by the central soft X-ray irradiation only. No other source of heating is necessary to produce the observed line ratios in this case. Follow up flickering mapping studies are suggested to probe the contribution of magnetic heating at the coupling region. The absolute phasing of the binary system has been updated using the narrow component of Balmer lines originated from the donor star.

MPD acknowledges the support by CNPq under grant #305725. We would like to thank F.J. Jablonski for providing us with scripts which facilitated the CCD photometry data reduction. We also thank L. Aceto and A. Bortoletto for their help.

# REFERENCES

Angelini, L., Osborne, J. P., Stella, L. 1990, MNRAS, 245, 652

Brainerd, J. J., Lamb, D. Q. 1985, ASSL, 113, 247

Araujo-Betancor, S., Gaensicke, B. T., Long, K. S., Beuermann, K., de Martino, D., Sion, E. M., Szkody, P. 2005, ApJ, 622, 589

Beuermann, K. & Schwope, A. D. 1994, ASPC, 56, 119

Cropper, M. 1988, MNRAS, 231, 597

Diaz, M. P., Steiner, J. E. 1994, A&A, 283, 508

Ferland, G. J., Korista, K. T., Verner, D. A., Ferguson, J. W., Kingdon, J. B., Verner, E. M. 1998, PASP, 110, 761

Ferrario, L., Wehrse, R. 1999, MNRAS, 310, 189

Gerke, J. R., Howell, S. B., Walter, F. M. 2006, PASP, 118, 678

Greeley, B. W., Blair, W. P., Long, K. S., Raymond, J. C. 1999, ApJ, 513, 491

Hamuy, M., Suntzeff, N. B., Heathcote, S. R., Walker, A. R., Gigoux, P., Phillips, M. M. 1994, PASP, 106, 566

Harrison, T. E., Osborne, H. L., Howell, S. B. 2005, AJ, 129, 2400

Howell, S. B., Cash, J., Mason, K. O., Herzog, A. E. 1999, AJ, 117, 1014

Kafka, S., Honeycutt, R. K. 2005, AJ, 130, 742

Liebert, J., Stockman, H. S., Williams, R. E., Tapia, S., Green, R. F., Rautenkranz, D., Ferguson, D. H., Szkody, P. 1982, ApJ, 256, 594

Marsh, T. R., Horne, K. 1988, MNRAS, 235, 269

Mennickent, R. E., Diaz, M. P., Arenas, J. 1999, A&A, 352, 167

Mukai, K., Charles, P. A. 1987, MNRAS, 226, 209

Ramsay, G., Cropper, M., Wu, K., Mason, K. O., Córdova, F. A., Priedhorsky, W. 2004, MNRAS,

    350, 1373

Rosenfeld, A. & Kak, A. C. 1982, Digital Picture Processing (New York: Academic Press)

Schwarz, R., Schwope, A. D., Staude, A., Remillard, R. A. 2005, A&A, 444, 213

Schwope, A. D. 2001, Lecture Notes in Physics, 573, 127

Schwope, A. D., Beuermann, K., Jordan, S., Thomas, H.-C. 1993, A&A, 278, 487

Schwope, A. D., Catalán, M. S., Beuermann, K., Metzner, A., Smith, R. C., Steeghs, D. 2000, MNRAS, 313, 533

Schwope, A. D., Naundorf, C. E., Thomas, H.-C., Beuermann, K. 1991, A&A, 244, 373

Shahbaz, T., Wood, J. H. 1996, MNRAS, 282, 362

Szkody, P., Liebert, J., Panek, R. J. 1985, ApJ, 293, 321

Warner, B. 1995, Cataclysmic Variable Stars (Cambridge: Cambridge Univ. Press)

Wickramasinghe, D. T., Cropper, M., Mason, K. O., Garlick, M. 1991, MNRAS, 250, 692

Williams, R. E., Ferguson, D. H. 1983, ASSL, 101, 97

FIGURE CAPTIONS

Figure 1. Differential I-band light-curves of MR Ser in high state. The upper panel shows the high state light curve taken during the first year of observations. The middle panel displays the data folded using the orbital ephemeris given in eq. 1 and a sinusoid fit to the observed magnitudes. The lower panel show the residuals of the sinusoid fit, phase-folded using $P_{ORB}$ / 2. In middle and lower panels the average magnitude is given with error bars representing the *RMS* value in each 0.03 phase interval, including both measurement and intrinsic variability dispersion. Estimated photometric uncertainty of unbinned data points is around 0.02 magnitudes.

Figure 2. Average continuum subtracted spectra of MR Ser in 8 phase bins. The effective phase of each bin is quoted on each spectrum. Tick marks on vertical axis are spaced by $5 \times 10^{-15}$ erg/s/cm$^2$/Å. The average spectra was shifted by $1.2 \times 10^{-14}$ erg/s/cm$^2$/Å for clarity. The spectral resolution is 1.9 Å (FWHM).

Figure 3. Trailed spectrograms of He II 4686 Å (left) and Hβ (right) line profiles. The continuum subtracted dataset used for the tomography analysis (see section 3.2) is shown here, binned into 0.03 phase intervals. The original spectral sampling of 0.45 Å / pix is retained. No intensity scaling or smoothing has been applied. The linear grayscale bar shown correspond to relative intensities ranging from 0.0 (white) to the maximum value (black) in both panels. The orbital cycle is repeated for clarity.

Figure 4. Doppler tomography images of MR Ser emission lines in high state. The velocity scale is the same for all panels and correspond to the intrinsic velocity scale at the orbital plane for an inclination of 45 degrees. The FWHM resolution of the maps are: 125 km/s (HeII, Hβ and Hγ), 140 km/s (HeI 4471 Å and HeI 4921 Å) and 180 km/s (CII 4267 Å). The pluses represent, from top to bottom, the secondary center of mass, the inner Lagrangian point (L1), the binary center of mass and the primary center. Curved lines show the free-particle path from L1 to the magnetic coupling radius $R_C \sim 15\ R_{WD}$ (for $M\text{dot} = 3\times10^{-10}\ M_{SUN}/\text{yr}$, $f = 0.001$ and $M_{WD} = 0.6\ M_{SUN}$) and the corresponding disk keplerian velocity at the stream. The straight line at the bottom left side of each graph represent the aproximate locus of the accretion column. Pluses on this line indicate a height of 10 and 5 $R_{WD}$. The normalized gray scale is the same for all panels. Contour lines in all panels are equally spaced from 10% (white) to 90% (dark gray) of maximum intensity in 10% steps. Maximum intensity values for HeII, Hβ, Hγ, HeI 4471 Å, HeI 4921 Å and CII 4267 Å tomograms are $1.5\times10^{-17}$, $1.6\times10^{-17}$, $1.4\times10^{-17}$, $4.6\times10^{-18}$, $3.1\times10^{-18}$ and $9.6\times10^{-19}$ erg/s/cm$^2$/(km/s)$^2$, respectively.

Figure 5. Distribution of the local HeII 4686 / Hβ emissivity ratio in velocity coordinates. The pluses and lines are as in figure 4. The FWHM resolution of the map is 150 km/s. The contour levels are equally spaced from 0.7 to 1.7 in 0.2 increments.

Figure 6. Accretion column mass density distribution along the accretion column as a function of the linear distance to the white dwarf assuming $B = 28$ MG and $M_1 = 0.5\ M_{SUN}$ (see text). The curves correspond from top to bottom to $M\text{dot} = 1.6\times10^{-9}$, $1.6\times10^{-10}$ and $1.6\times10^{-11}\ M_{SUN}/\text{yr}$.

Figure 7. HeII 4686 / Hβ model ratios and gas temperatures (electron temperature weighted by the electron density) for an accretion column/stream illuminated from outside by the soft X-ray source at white dwarf surface with $L_{BB} = 8\times10^{32}$ erg.s$^{-1}$ and $kT_{BB} = 12$ eV. $\dot{M}$ shown is $3\times10^{-10}$ $M_{SUN}$/yr. Typical Hydrogen densities for solar abundances are $N_H \sim 10^{13}$ cm$^{-3}$. Values for $r < 7$ $R_{WD}$ are uncertain due to the higher densities expected in the column (see text).

| TABLE 1 | | | | |
|---|---|---|---|---|
| JOURNAL OF OBSERVATIONS | | | | |
| Date | HJD (start) | Number of Exposures | Telescope | Instrument |
| May 27, 2006 | 53882.72 | 7 | 1.6m | CassSpec |
| May 28, 2006 | 53883.50 | 44 | 1.6m | CassSpec |
| May 29, 2006 | 53884.51 | 47 | 1.6m | CassSpec |
| May 30, 2006 | 53885.51 | 45 | 1.6m | CassSpec |
| May 30, 2006 | 53885.54 | 716 | 0.6m | CCD Camera |
| May 31, 2006 | 53886.55 | 37 | 1.6m | CassSpec |
| Jun 01, 2006 | 53887.57 | 13 | 1.6m | CassSpec |
| May 22, 2007 | 54242.68 | 10 | 1.6m | CassSpec |
| Jun 15, 2007 | 54267.47 | 17 | 1.6m | CassSpec |
| Jun 17, 2007 | 54268.51 | 31 | 1.6m | CassSpec |
| Jun 18, 2007 | 54269.50 | 20 | 1.6m | CassSpec |

| TABLE 2 | | |
|---|---|---|
| TOMOGRAM AND MODEL LINE FLUXES | | |
| Line | Observed flux[a] ($H\beta = 1$) | Model #109[b] ($H\beta = 1$) |
| HeII 4686 | 1.49 | 1.49 |
| HeI 4922 | 0.21 | 0.21 |
| HeI 4471 | 0.35 | 0.34 |
| $H\beta$[c] | $5.0 \times 10^{-15}$ | $3.9 \times 10^{-15}$ |
| $H\gamma$ | 0.90 | 0.93 |

[a] Average fluxes at $r \approx R_c$.
[b] Model fluxes for $E(B-V) = 0$ and $d = 150$ pc.
[c] Flux in erg.cm$^{-2}$.s$^{-1}$.

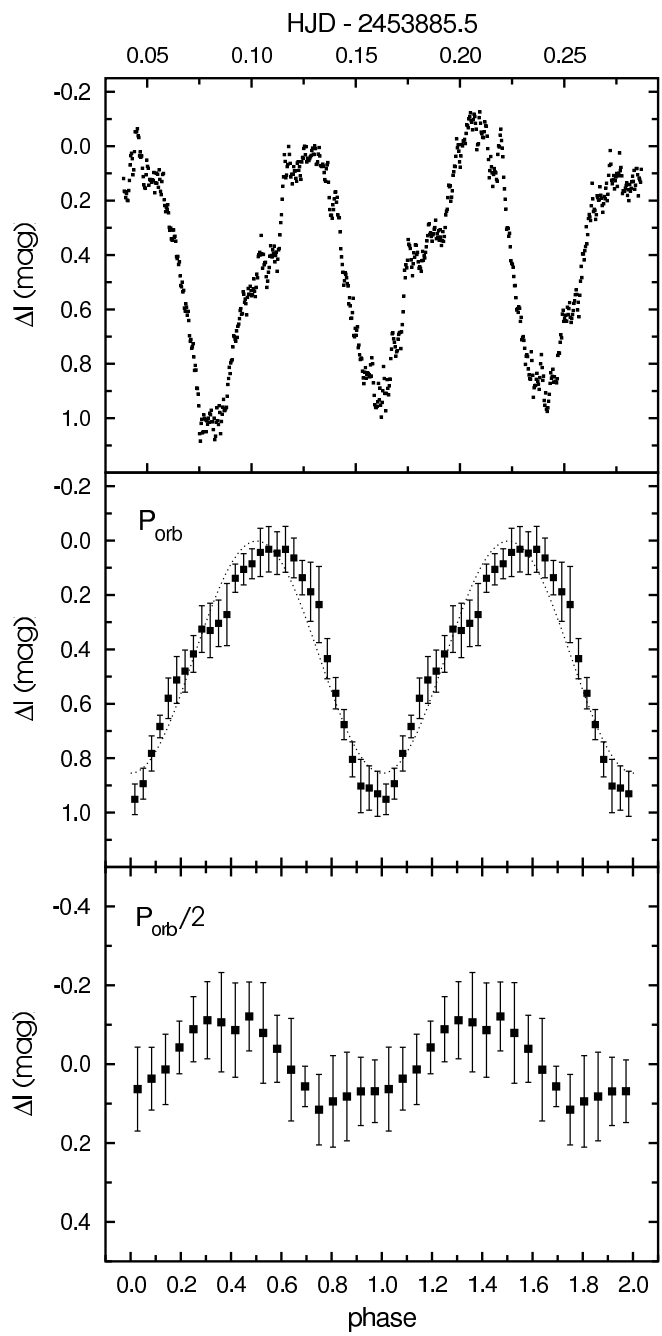

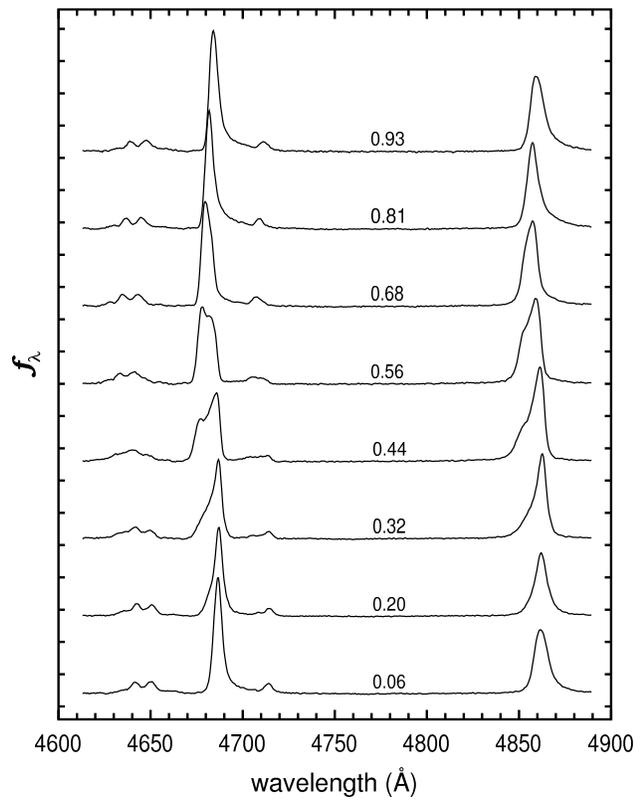

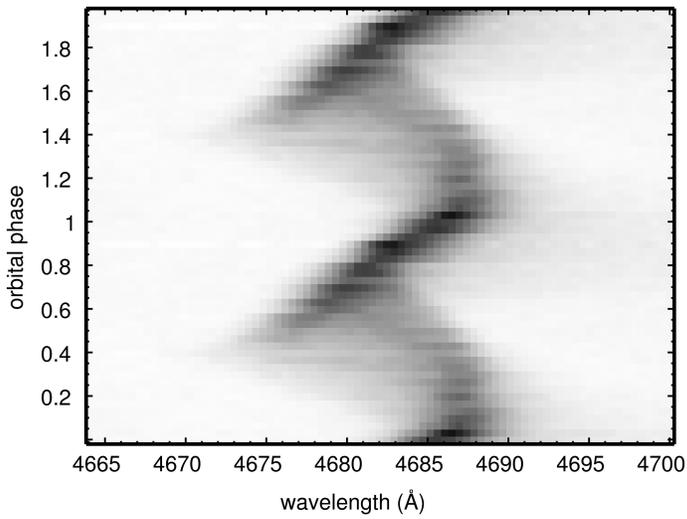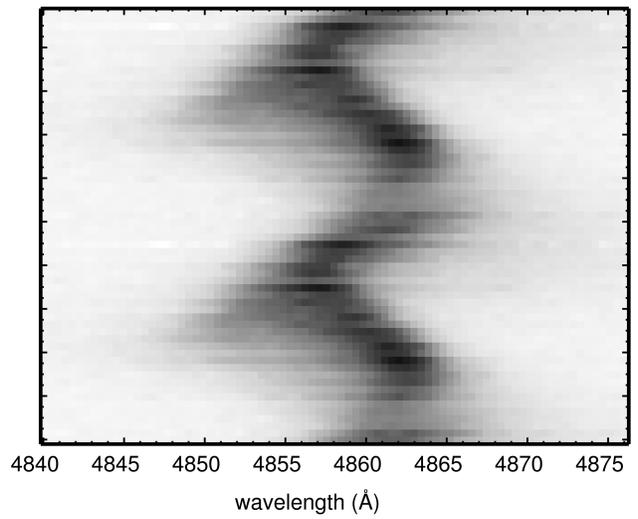

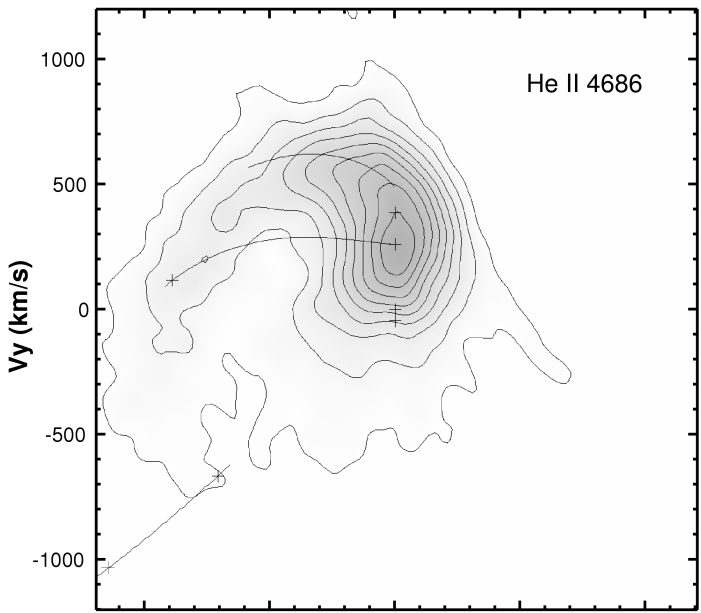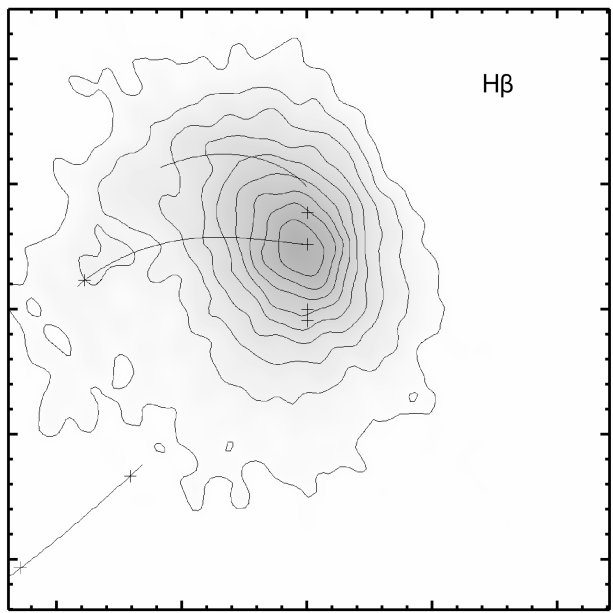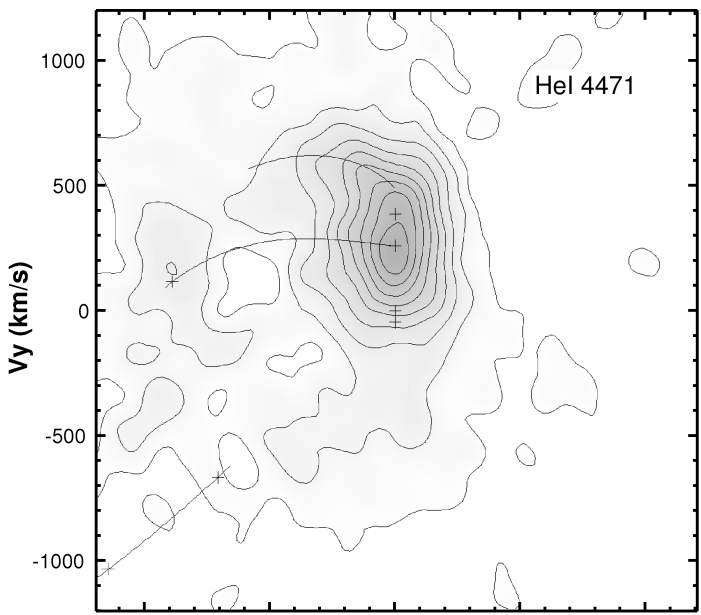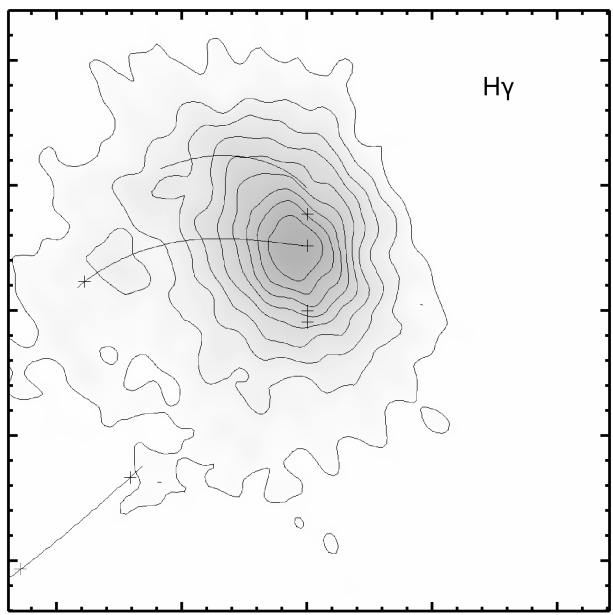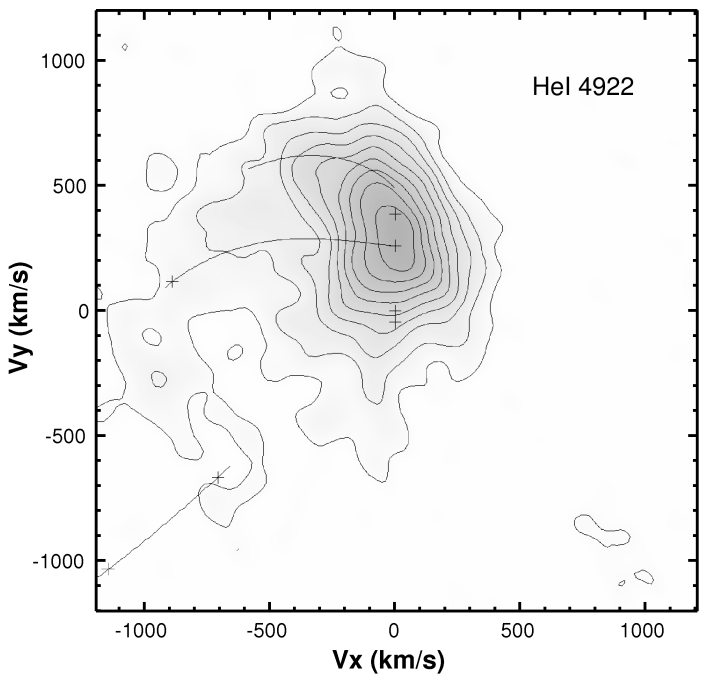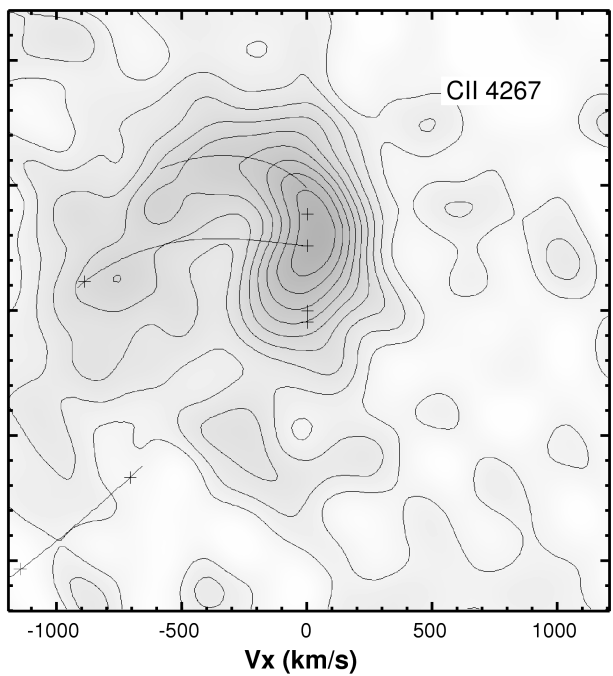

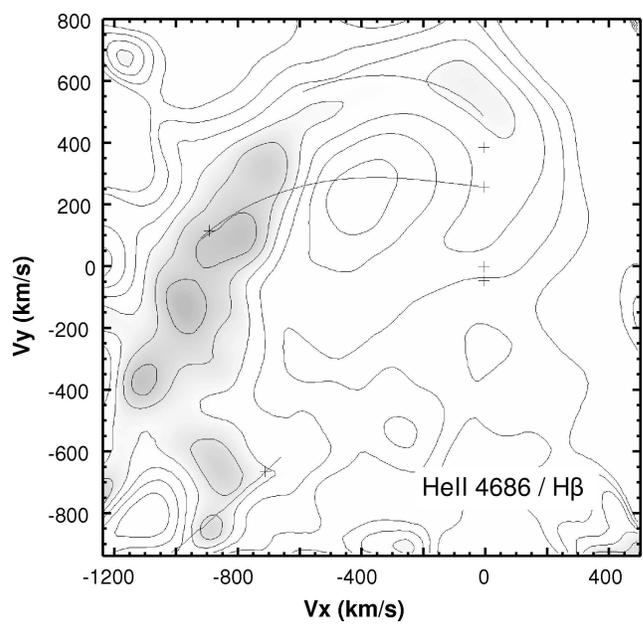

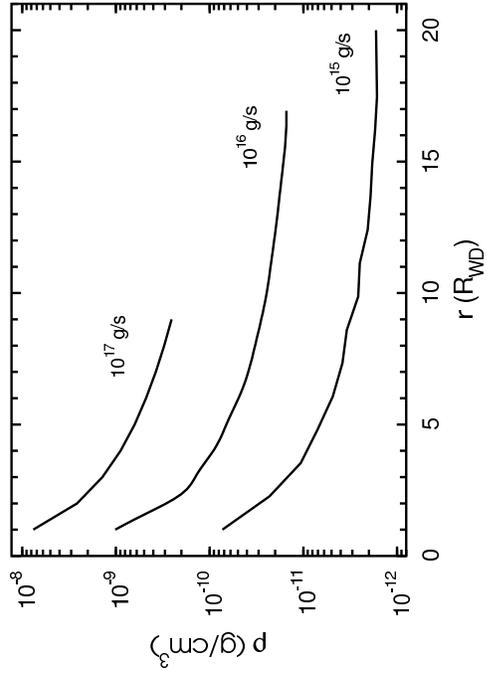

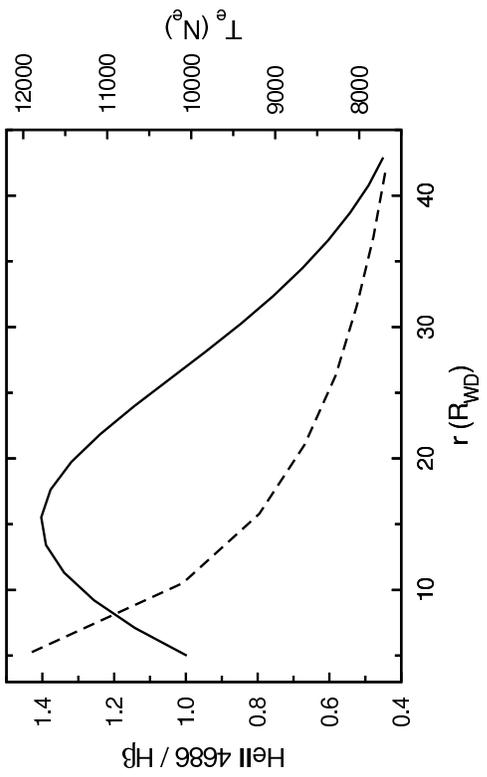